\documentclass[aps,prx,reprint,amsmath,amssymb,superscriptaddress]{revtex4-2}
\usepackage[normalem]{ulem}
\usepackage{dsfont}
\usepackage{tabu}

\newcommand{\abs}[1]{\left|#1\right|}

\usepackage{graphicx}
\usepackage[colorlinks]{hyperref}

\usepackage{soul}

\usepackage{xcolor}

\begin{document}

\title{Scalable multiphoton quantum metrology with neither pre- nor post-selected measurements}

\author{Chenglong You}
\affiliation{Quantum Photonics Laboratory, Department of Physics \& Astronomy, Louisiana State University, Baton Rouge, LA 70803, USA}

\author{Mingyuan Hong}
\affiliation{Quantum Photonics Laboratory, Department of Physics \& Astronomy, Louisiana State University, Baton Rouge, LA 70803, USA}

\author{Peter Bierhorst}
\affiliation{Mathematics Department, University of New Orleans, New Orleans, Louisiana 70148, USA}

\author{Adriana E. Lita}
\affiliation{National Institute of Standards and Technology, 325 Broadway, Boulder Colorado 80305, USA}

\author{Scott Glancy}
\affiliation{National Institute of Standards and Technology, 325 Broadway, Boulder Colorado 80305, USA}

\author{Steve Kolthammer}
\affiliation{QOLS, Blackett Laboratory, Imperial College London, London SW7 2AZ, United Kingdom}

\author{Emanuel Knill}
\affiliation{National Institute of Standards and Technology, 325 Broadway, Boulder Colorado 80305, USA}
\affiliation{Center for Theory of Quantum Matter, University of Colorado, Boulder, Colorado 80309, USA}

\author{Sae Woo Nam}
\affiliation{National Institute of Standards and Technology, 325 Broadway, Boulder Colorado 80305, USA}

\author{Richard P. Mirin}
\affiliation{National Institute of Standards and Technology, 325 Broadway, Boulder Colorado 80305, USA}

\author{Omar S. Maga\~na-Loaiza}
\email[]{maganaloaiza@lsu.edu}
\affiliation{Quantum Photonics Laboratory, Department of Physics \& Astronomy, Louisiana State University, Baton Rouge, LA 70803, USA}

\author{Thomas Gerrits}
\affiliation{National Institute of Standards and Technology, 100 Bureau Drive, Gaithersburg, MD 20899, USA}

\date{\today}

\begin{abstract}
The quantum statistical fluctuations of the electromagnetic field establish a limit, known as the shot-noise limit, on the sensitivity of optical measurements performed with classical technologies. However, quantum technologies are not constrained by this shot-noise limit. In this regard, the possibility of using every photon produced by quantum sources of light to estimate small physical parameters, beyond the shot-noise limit, constitutes one of the main goals of quantum optics. Here we experimentally demonstrate a scalable protocol for quantum-enhanced optical phase estimation across a broad range of phases, with neither pre- nor post-selected measurements. This is achieved through the efficient design of a source of spontaneous parametric down-conversion in combination with photon-number-resolving detection. The robustness of two-mode squeezed vacuum states against loss allows us to outperform schemes based on N00N states, in which the loss of a single photon is enough to remove all phase information from a quantum state. In contrast to other schemes that rely on N00N states or conditional measurements, the sensitivity of our technique could be improved through the generation and detection of high-order photon pairs. This unique feature of our protocol makes it scalable. Our work is important for quantum technologies that rely on multiphoton interference such as quantum imaging, boson sampling and quantum networks.

\end{abstract}

\maketitle

The quantum theory of electromagnetic radiation predicts statistical fluctuations for diverse sources of light \cite{Glauber:1963, Dodonov:2002,gerry2005introductory,agarwalbook}. Indeed, the underlying amplitude and phase fluctuations of photons establish a quantum limit for the uncertainty of optical measurements made using classical states of light \cite{caves:1981}. These fundamental properties of light have enabled scientists to identify the shot-noise limit (SNL) as the classical limit for precision measurements \cite{Jarzyna:12,Lang:13,Takeoka:17}. Remarkably, states of light characterized by nonclassical statistical properties can, in principle, surpass the SNL \cite{caves:1981,Giovannetti:04,Giovannetti:11,Krischek:11,Datta:11,Motes:15,demkowicz2015,You:19,Lang2}. This possibility has triggered an enormous interest in the experimental demonstration of quantum-enhanced measurements \cite{Walmsley:10,Yonezawa1514,Walmsley:14,goda2008quantum}. Unfortunately, efficient and scalable schemes for quantum metrology have remained elusive due to stringent technical requirements.  Nevertheless, the possibility of estimating small physical parameters with greater sensitivity has important implications for diverse quantum technologies ranging from quantum communication to quantum information processing \cite{Giovannetti:06,Fabio:06,LIGO:11, Omar2019Review, You2020Review, YouAPR2020}.

\begin{figure*}[!htbp]
  \centering
  \includegraphics[width=0.95\textwidth]{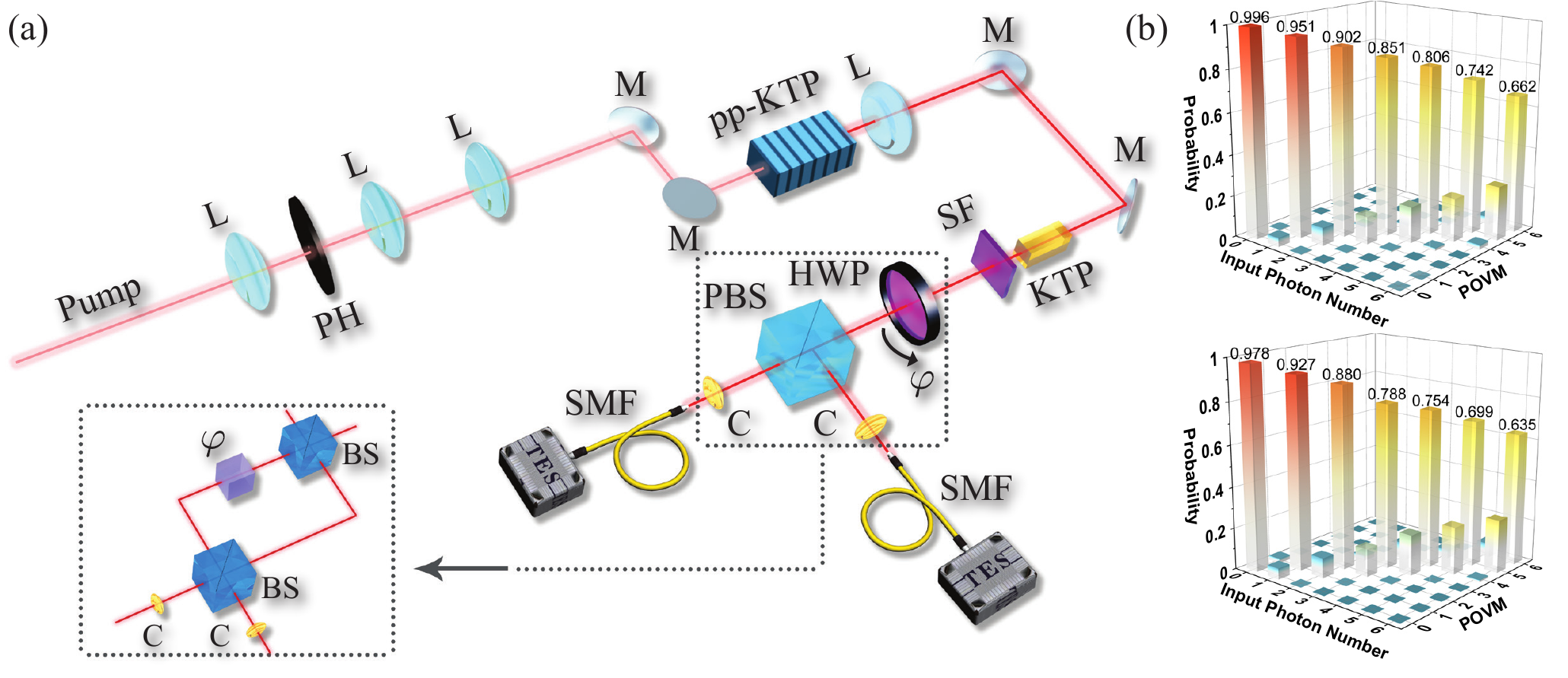}
  \caption{Experimental setup used to demonstrate multiphoton quantum-enhanced phase estimation. (a) We generate two-mode squeezed vacuum (TMSV) states through a type-II parametric down-conversion process. This is achieved by pumping a periodically poled potassium titanyl phosphate (ppKTP) crystal with a femtosecond Ti:Sapphire laser at 775 nm, cavity dumped with a repetition rate of 76 MHz, which is then pulse-picked at a repetition rate of 229.166 kHz. We use collimation and collection lenses in combination with a 20 $\mu$m pinhole (PH) to filter the spatial profile of the pump beam. The down-converted photons are spectrally filtered by an antireflection-coated piece of silicon (SF) and then injected into a common-path Mach-Zehnder interferometer. Here, two paths of the Mach-Zehnder are replaced with two polarization modes H and V, so that the common-path interferometer implements the same transformation as the traditional Mach-Zehnder interferometer that is shown in the inset. The emerging photons are then coupled into single mode fibers (SMFs) which direct the photons to two transition edge sensors (TESs) with photon number resolution. (b) Quantum detector tomography of our two TESs, we show the reconstructed diagonal positive-operator valued measures (POVMs) for up to 6 photons. L: Lens; M: Mirror; HWP: Half Wave Plate; C: Collimating Lens; BS: Beam Splitter. }
\label{schematic}
\end{figure*}

Over the past three decades, there has been a strong impetus to employ quantum multiphoton states, as well as diverse measurement schemes, to demonstrate estimation of phase shifts with sensitivities that surpass the SNL \cite{slusssarenko:17, Sciarrino20review}. Hitherto, N00N states, an example of multiphoton path-entangled states, have been extensively used to demonstrate super-resolution in quantum metrology \cite{DowlingPRL2000, Dowling:08, Anisimov:10, Nagata:07, Su:17, Walther:04, Afek879, Sciarrino20review}.  Despite the relevance that these quantum states of light have for quantum metrology, their generation is extremely difficult, particularly when the N00N state involves more than two photons \cite{Sciarrino20review}. These challenges have motivated the exploration of conditional schemes that rely on pre- or post-selective measurements \cite{Walther:04, Mitchell2004, Thekkadath20, Nagata:07, Dowling09, Afek879, Walmsley11PRL, Obrien:16, Su:17}. 
In experiments with ``pre-selection'', researchers have increased the resolution of phase measurements by conditioning experiments on the observation of specific photon detections that herald the presence of a desired quantum state before the state interacts with the sample to be measured \cite{Nagata:07, Su:17, Thekkadath20}. Conversely, implementations of optical phase estimation based on ``post-selection'' have been demonstrated by conditioning the outcomes of experiments on specific detection events after the probe state has passed through the sample \cite{Walther:04, Dowling09, Mitchell2004, Afek879, Walmsley11PRL, Obrien:16, liu2020distributed}. Unfortunately, both protocols discard photons that represent valuable resources for interrogation of phase elements, and in practice pre- and post-selected schemes often succeed with low probability. Thus far, all demonstrations of quantum metrology for phase estimation have relied on either pre- or post-selected measurements or have lacked photon-number resolution \cite{Walther:04,Mitchell2004, Walmsley11PRL, Nagata:07, Dowling09, Afek879, Obrien:16, slusssarenko:17, Su:17, Sciarrino20review, Thekkadath20, liu2020distributed}. Nevertheless, theoretical studies suggest that counting of resources only when pre- or post-selection is successful may lead to enhanced sensitivities beyond the SNL \cite{Davidovich15PRA, ZhangWalmsley15PRL, lloyd2020_post}. 

In 2017, Slussarenko and co-workers demonstrated the first N00N-state based quantum metrology protocol that did not rely on post-selection \cite{slusssarenko:17}. That protocol relied on the generation of a high-fidelity N00N state with N=2 photons. Sharing similarities with other schemes that rely on N00N states, the loss of a single photon in this protocol removes all phase information encoded in the N00N state \cite{Walmsley11PRL}. This vulnerability results in higher uncertainties when using N00N states than when using two-mode squeezed vacuum (TMSV) state for the estimation of optical phase shifts \cite{Walmsley11PRL,Wang:13,Knott14PRA}. In addition, higher-order photon generation in spontaneous parametric down-conversion (SPDC) induces errors in protocols that rely on specific photon events to generate perfect N00N states \cite{slusssarenko:17}. Despite the importance of conditional measurements for quantum technologies \cite{Omar:19}, the possibility of using every photon produced by quantum sources to perform quantum parameter estimation constitutes one of the main goals of quantum optics. Consequently, the first demonstration of scalable multiphoton quantum metrology using photon counting detectors with neither pre- nor post-selected measurements represents an important progression in the field of quantum photonics. Here, we demonstrate a scalable scheme that utilizes all the detected photons from a two-mode squeezed vacuum state to perform unconditional quantum-enhanced phase estimation with an overall system efficiency of 80$\%$. We define a scalable optical phase measurement protocol to be one whose Fisher information per photon (that interacts with the sample being measured) exceeds the standard quantum limit and increases with the mean photon number. This technique relies on an efficient source of SPDC and transition edge sensors that enable the implementation of photon-number-resolving detection. We demonstrate multiphoton quantum enhancement in the estimation of 62\% of the complete range of optical phases, without conditional measurements.

\begin{figure*}[!htbp]
  \centering
  \includegraphics[width=0.95\textwidth]{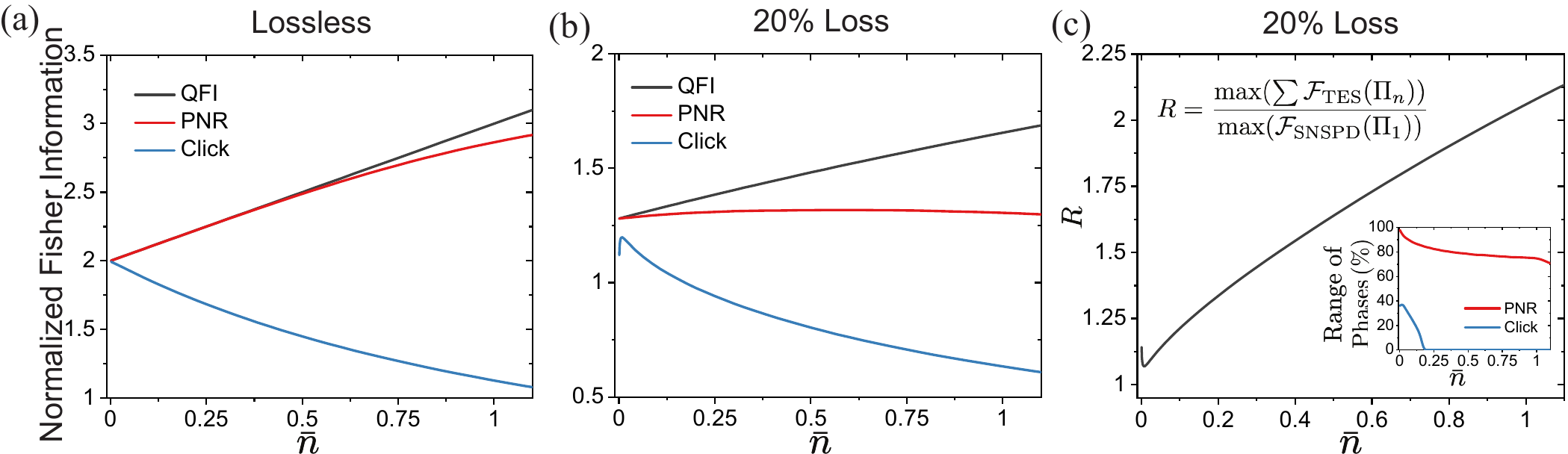}
  \caption{Scalability of multiphoton quantum metrology quantified through theoretically calculated Fisher information. In (a) and (b) we plot the dependence of Fisher information for different conditions of loss for a quantum metrology protocol relying on photon-number-resolving (PNR) and click detection. The Fisher information is normalized with respect to the mean photon number interacting with the sample being measured. In this case, we have selected phase angles that lead to the best phase estimation using PNR (red) and click (blue) detection. In black we also show the normalized quantum Fisher information (QFI), which is the maximum Fisher information achievable with any detection scheme. In (c) we show our estimation of the ratio of the classical Fisher information for our multiphoton quantum metrology protocol with respect to protocols based on click detection, maximized over phases. Additionally, the inset in (c) shows the percentage of phases that can be estimated with sub-SNL uncertainty for both protocols. Here, we assumed that the TES detectors can resolve up to 10-photon events under the loss conditions of our experiment.}
  \label{compare}
\end{figure*}

Our experimental setup is depicted in Fig. \ref{schematic}(a). We use a Ti:Sapphire laser that is spatially filtered by a 20 $\mu$m pinhole and then focused to pump a type-II periodically poled potassium titanyl phosphate (ppKTP) crystal with a length of 2 mm and a poling period of 46.1 $\mu$m. We produce TMSV state with orthogonal polarization H and V at 1550~nm through a process of spontaneous parametric down-conversion. Ideally, the TMSV state is described by $| z \rangle=\sqrt{1-|z|^{2}} \sum_{n=0}^{\infty} z^{n} | n \rangle_{\text{s}} | n \rangle_{\text{i}}$, where $n$ denotes the number of photons in the signal (s) and idler (i) modes, and $z$ represents the squeezing parameter. The parameter $z$  depends on the nonlinear properties of the crystal, as well as the pump mode and pump power. The filtered signal and idler photons pass through a rotated KTP crystal that compensates for temporal differences between signal and idler photons. Our protocol for phase estimation uses a common-path Mach-Zehnder interferometer (MZI). Here, the two paths of a conventional MZI are replaced with two polarization modes H and V. The phase $\phi$ that we estimate in our experiment is controlled by a half-wave plate (HWP). Finally, signal and idler photons interfere at a polarizing beam-splitter (PBS). The transmitted and reflected photons are then collected by single-mode fibers, and detected by superconducting transition edge sensors (TESs) with photon-number resolution.

In order to illustrate the fundamental differences between our scalable protocol for multiphoton quantum metrology and previous schemes based on click detection, we introduce operators that describe the measurements performed by a TES and a superconducting nanowire single photon detector (SNSPD). The latter is used to implement click detection. First, we define the ideal TESs' positive-operator valued measure (POVM) corresponding to the detection of $n$ photons as $\Pi_{n}=|n\rangle\langle n|$. Similarly, we describe the ideal click detection operator associated to a SNSPD as $\Pi_0=|0\rangle\langle 0|$ and $\Pi_\text{c}=\mathbb{I}-\Pi_0$. Here, $\Pi_0$ represents a vacuum detection, and $\Pi_\text{c}$ represents a click detection. As shown below, our TES-based protocol is more sensitive than the SNSPD-based protocol and scales to larger photon numbers \cite{Obrien:16, slusssarenko:17, liu2020distributed}. This particular feature makes our protocol scalable.

We demonstrate this feature by analyzing the Fisher information obtained through the implementation of photon-number-resolving and click detection. For this purpose, we maximized the Fisher information over phase angles that lead to the best phase sensitivity using both detection schemes, while keeping mean photon number, losses and the photon-number resolution of our TESs fixed. Here, we have assumed that our TESs can resolve up to 10-photons. However, it is worth mentioning that TESs can efficiently resolve up to 100 photons at 1550 nm with high fidelities \cite{Levine:14, Gerrits:12}. In Figs. \ref{compare} (a) and (b), we plot the quantum Fisher information (black line) as well as the classical Fisher information for photon-number-resolving (red) and click (blue) detection. In this case, Fisher information is normalized with respect to the mean photon number interacting with the sampling being measured. This metric enables one to quantify the Fisher information gained per photon, as well as the performance of the protocols with respect to the SNL. These plots show that the Fisher information obtained through photon-number-resolving detection is always higher than click detection. Interestingly, in the lossless region, the performance of our metrology protocol approaches the quantum Fisher information. This characteristic feature differentiates our multiphoton quantum metrology protocol from previous schemes \cite{Obrien:16,slusssarenko:17}.

\begin{figure*}[!htbp]
  \centering
  \includegraphics[width=0.95\textwidth]{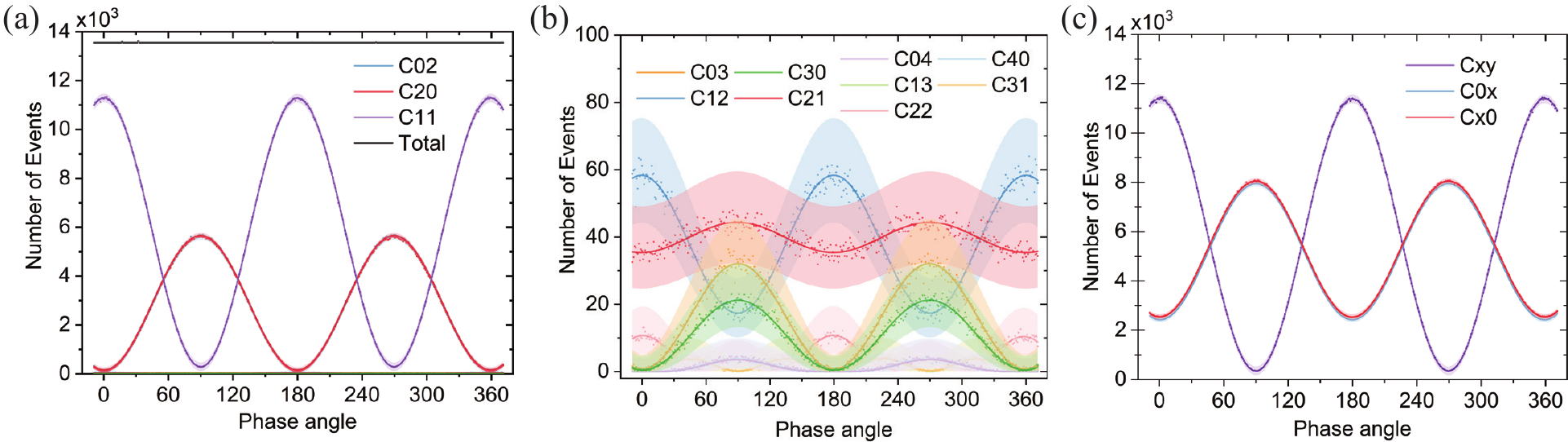}
   \caption{Experimental measurement of multiphoton interference events in our protocol for quantum metrology. The detection events produced by multiphoton interference are shown in (a) and (b). We use ``C$mn$" to label the detection of $m$ photons in TES1 and $n$ photons in TES2. In (a), the C20 data overlaps C02. In (b) we expand the vertical axis to show the higher photon-number detection events, which are not clearly visible at the bottom of (a). Furthermore, in (c) we show how the data of (a) and (b) would have appeared if we used click detectors. Here ``Cxy" shows events with at least 1 photon detected in both modes, ``C0x" shows events with no photons detected in signal mode and at least one detected in idler mode, and ``Cx0" shows events with at least one photon detected in signal mode and no photons detected in idler mode. We use a theoretical model together with our photon-number-resolving measurements to obtain the traces in (c). The data points (dots) in (a) and (b) are generated from averaging nine repeated experimental trials, and each line represents the fit to the nine experimental trial. Each realization of the experiment took approximately 2 hours to gather. All shaded areas correspond to the 95\% confidence bounds for the fitted number of events.}
\label{events}
\end{figure*}

Moreover, in Fig. \ref{compare} (c), we report an estimation of the ratio of the maximum classical Fisher information of our experiment $\max{(\mathcal{F}_{\text{TES}})}$ with respect to $\max{(\mathcal{F}_{\text{SNSPD}})}$. The results in Fig. \ref{compare} (c) indicate that our protocol outperforms metrology schemes based on click detection  \cite{Obrien:16,slusssarenko:17}. This advantage scales with the mean photon number of the interrogating photons. Additionally, the inset plot shows the percentage of phases that can be estimated with sub-SNL performance for click and photon-number-resolved detection. Fig. \ref{compare} demonstrates the superior robustness of our protocol against losses with respect to those relying on click detection schemes. Naturally, losses and the limited resolution of the TES diminishes the sensitivity of quantum protocols for multiphoton metrology. Nevertheless, the latter limitation can be alleviated by multiplexing multiple detectors \cite{Harder16PRL}.

In order to characterize the performance of our experimental setup, we use calibrated SNSPDs. These detectors are used to measure the Hong-Ou-Mandel visibility of $\mathcal{V}=99.36(5)\%$ for our common-path interferometer. Note that for the HOM measurement, we rotate the HWP by 45 degrees at zero delay. We also determine the overall detection efficiencies (source to SNSPD) of the system. This in turn allows us to estimate the overall source-to-fiber efficiencies for each of the arms. In our experiment, we implement photon-number-resolving detection with TESs that were characterized through quantum detector tomography \cite{Lundeen:09}. We estimate the efficiencies of our TESs from the tomography data shown in Fig. \ref{schematic}(b). The details of the tomography method can be found in the Supplemental Material. Furthermore, we estimate the overall system detection efficiencies (source to TES) for each of the outputs of our experiment. All the efficiencies are listed in Table \ref{tab}.  

\small
\begin{table}[!htbp]
\centering
\caption{\label{tab}Experimental efficiencies of our setup.}
{\def\arraystretch{1}\tabcolsep=8pt
\begin{tabular}{ccc}
\hline
Efficiency   & Arm1      & Arm2      \\ \hline
SNSPD & 95.9(5)\%   & 92.3(5)\%   \\
Overall (SNSPD)   & 83.42(7)\%  & 79.11(7)\%  \\
Source to Fiber   & 87.0(9)\%   & 85.7(9)\%  \\
TES   & 92.6(1.0)\% & 95.1(1.0)\% \\
Overall (TES) & 80.5(1.2)\% & 81.5(1.2)\% \\ \hline
\end{tabular}
}
\end{table}
\normalsize

As shown in Fig. \ref{events}(a) and (b), the TESs enable us to probe all the possible multiphoton interference events, up to 4 photons. The black dotted line in Fig. \ref{events}(a) shows that the total number of detected photons does not depend on the phase. We omit the vacuum events and single-photon events in the plot, which are nearly constant for varying phase angles and therefore do not contribute significantly to the ability to detect changes in this parameter (See Supplemental Material). This means that these events do not contribute to the sensitivity of our phase estimation measurements. Moreover, in Fig. \ref{events}(b) we plot the traces produced by three- and four-particle interference. The photon-number resolution of our protocol enables the identification of the complex multiparticle interactions that take place in our experiment. Every photon count conveys relevant information regarding the phase that we aim to estimate. This information is not available in quantum metrology protocols that rely on click detectors, which can distinguish zero from one or more photons but otherwise lack photon number resolution. Furthermore, our scheme enables the estimation of a broad range of phases with sensitivities that surpass the SNL. To compare our results to those that would be obtained without photon number resolving detectors, we reconstruct interference structures obtained with click detectors in Fig. \ref{events}(c).

\begin{figure}[!htbp]
  \centering
  \includegraphics[width=0.47\textwidth]{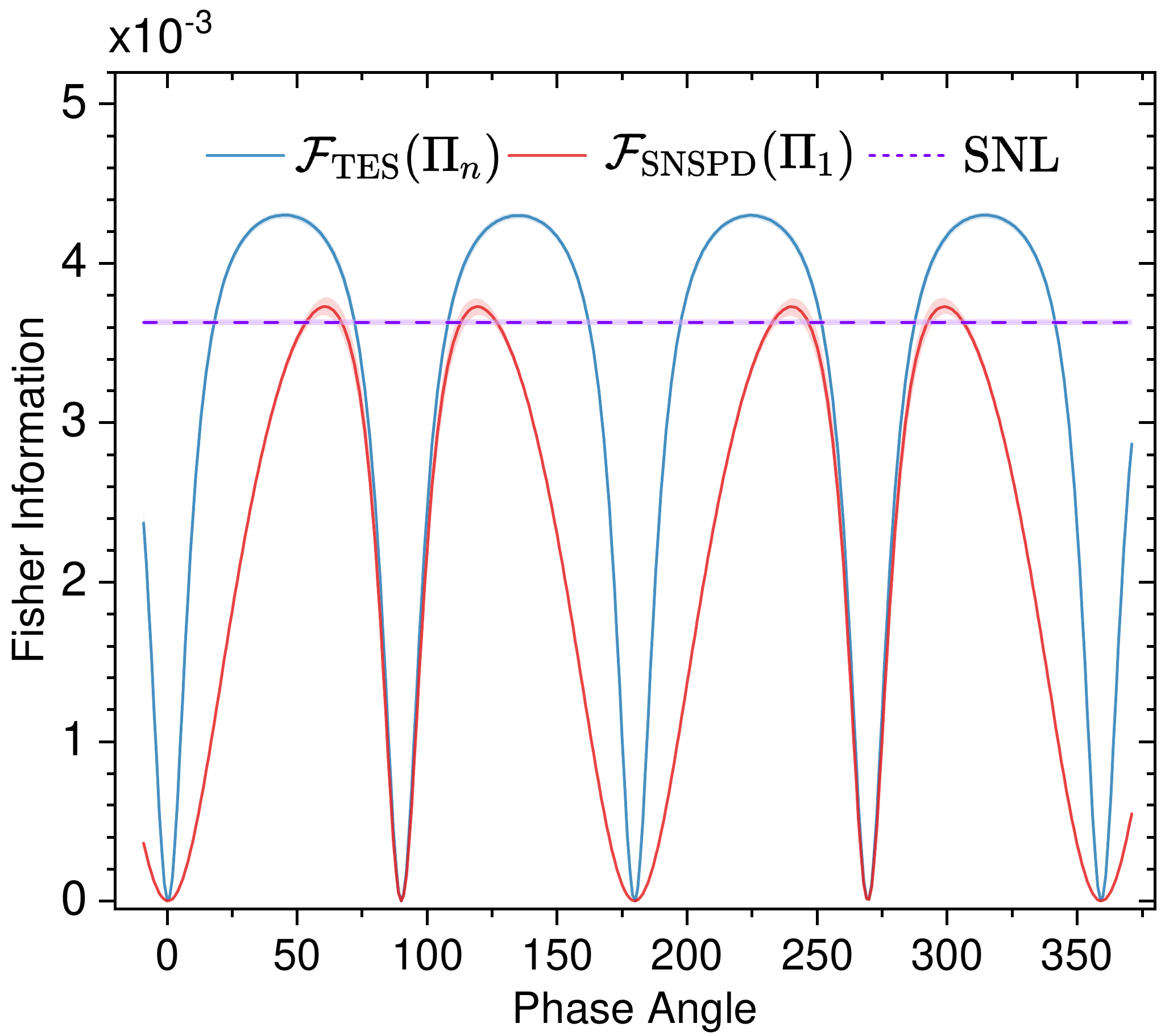}
  \caption{Experimental Fisher information obtained by using the photon events shown in Fig. \ref{events}. The blue (red) line represents the Fisher information of our setup with (without) photon-number-resolving detection. The purple dashed line represents the SNL of our experiment. This limit is established by fitting our experimental data to the theoretical model of our experiment. Here, the SNL is calculated by estimating the number of photons generated through the process of spontaneous parametric down-conversion. The faint shaded areas correspond to the 95\% confidence bounds. The 95\% confidence bounds for the Fisher information is calculated based on a bootstrapping method using the data shown in Fig. \ref{events}. We use the uncertainty of the fitted parameters in the theoretical model of our experiment to calculate the 95\% confidence bound for the SNL.}
  \label{fisher}
\end{figure}

To quantify the performance of our metrology device, we use the classical Fisher information $\mathcal{F}=\sum_{i}\left(\frac{\partial \ln p_{i}}{\partial \phi}\right)^{2} p_{i}$, where the probability $p_i$ includes all possible detection results for an experimental trial, i.e. $i\in\{00,01,10,11,...\}$, so $\mathcal{F}$ quantifies the information gained in each trial. Each $p_i$ is estimated using the data given in Fig. \ref{events}. Probabilities that do not vary with $\phi$, such as the probability $p_{00}$ of a vacuum event, contribute zero for the corresponding term in the sum defining $\mathcal{F}$. As shown in Fig. \ref{fisher}, we have calculated the classical Fisher information $\mathcal{F}_{\text{TES}}$ and $\mathcal{F}_{\text{SNSPD}}$. We also calculate the SNL given by $\mathcal{F}_{\text{SNL}}=\bar{n}$, the average number of photons generated per experimental trial, derived from our experimental results. The average photon number $\bar{n}= 3.631(14) \times 10^{-3}$, is estimated by fitting our experimental data to the theoretical model of our setup. Such an approach takes all losses into account, including the mode mismatch at the generation of entangled photons, fiber-coupling loss as well as the loss of our TESs. The details of the calculation can be found in the Supplemental Material. The blue line represents the Fisher information estimated for a detection scheme with photon-number resolution. As a comparison, we plot the Fisher information using the click detector data presented in Fig. \ref{events} (c), which shows lower Fisher information at all angles. These results confirm the fundamental difference between the implementation of the operator
$\Pi_{n}$ and $\Pi_{\text{c}}$ as discussed in Ref. \cite{Anisimov:10}. It is also worth noticing that tha click detector measurement barely reaches the SNL represented by the purple dashed line, whereas the measurement with TESs surpasses the SNL for an impressively broad range of 62\% of the phase angles. It is worth highlighting the robustness of our protocol against losses. Remarkably, our scheme for quantum metrology allows for phase estimation with sensitivities that surpass the SNL, even in dissipative environments with up to 28\% loss, meaning that in our specific experimental realization, inserting a sample with up to about 11\% loss would still have surpassed the SNL at a phase angle of $\phi=\pi/4$. This prediction is obtained based on our theoretical model outlined in the Supplementary Material. These results demonstrate the superior performance of our metrology scheme utilizing photon-number-resolved detection. In contrast to other schemes that rely on N00N states or conditional measurements, the sensitivity of our technique is improved through the generation and detection of high-order photon pairs. This unique feature of our protocol makes it scalable.  
 
Over the past three decades, scientists and engineers have developed technology and explored novel quantum states of light, as well as quantum measurement schemes, with the aim to demonstrate scalable and unconditional quantum protocols. For the first time, we demonstrated the possibility of using photon-number resolution detectors with quantum sources of light to perform scalable estimation of optical phase shifts with sensitivities that surpass the SNL over a wide range. The broad-peaked Fisher information in our metrology experiment enables one to exceed the shot-noise limit for almost 62\% of the phase space even in the presence of losses. This feature of our technique is important for quantum-enhanced phase measurement applications such as quantum imaging \cite{lugiatoimaging,Brida:10}. Our multiphoton quantum metrology protocol increases Fisher information per photon compared to the click detection protocol. This possibility has shown important implications for measuring and analyzing sensitive properties of delicate samples \cite{PhysRevApplied.15.044012,PhysRevLett.123.023601}. Furthermore, our technique is relevant for other quantum technologies that rely on nonclassical multiphoton interference, such as boson sampling and quantum networks \cite{Fabio:06,Walther:13,Pan:17,Spring798}.

\begin{acknowledgments}
C.Y., M.H., and O.S.M-L. acknowledge funding from the U.S. Department of Energy, Office of Basic Energy Sciences, Division of Materials Sciences and Engineering under Award DE-SC0021069. T.G., C.Y. and O.S.M-L. thank Animesh Datta, Peter Humphreys, Oliver Slattery and Alan Migdall for helpful discussions. We dedicate this article to late Prof. J. P. Dowling. He motivated scientists all around the world to work on the fascinating field of quantum metrology. 
\end{acknowledgments}

\section*{COMPETING INTERESTS}
The authors declare that they have no competing interests.

\section*{DATA AVAILABILITY}
The data sets generated and/or analyzed during this study are available from the corresponding author or last author on reasonable request.

\bibliography{refs.bib}

\onecolumngrid
\appendix

\section*{SUPPLEMENTAL MATERIAL}

\subsection*{Detectors tomography}

We characterize our TESs by performing quantum detector tomography. The purpose of this technique is to characterize the TES response given a photon number state in the input. In this context, the detector response is described by the detector's positive-operator valued measure (POVM). Since it is not possible to generate arbitrary Fock states on demand, we use varying weak coherent states and employ a tomographic reconstruction of the POVMs \cite{Lundeen:09}. For both TESs, we assume no phase dependence; therefore the POVM elements are diagonal in the Fock basis.  The POVM element corresponding to detecting $n$ photons is $\{\pi_{n}\}$, and the $k^\text{th}$ diagonal entry of $\pi_n$ is $\pi_{n}(k)=\theta_{k}^{(n)}|k\rangle\langle k|$, where $|k\rangle$ is the $k$-photon Fock vector.
 We probe the TESs with a set of $m$ coherent-state probes with different amplitudes $\alpha_{m}$. A coherent state with amplitude $\alpha_{m}$ has probability $C_{m,k}=\left|\alpha_{m}\right|^{2 k} \exp \left(-\left|\alpha_{m}\right|^{2}\right)/k!$ to contain $k$ photons. Here the Hilbert space needs to be truncated at some finite photon number, and for our purpose we choose to truncate at $k=9$. By constructing the POVM set matrix $\Pi$ as $\Pi_{k,n}=\theta_{k}^{(n)}$, we can write the measured statistics matrix $R$ as a matrix equation $R = C\Pi$. Therefore, the matrix element $R_{m,n}$ represents the probability to observe $n$ photons when a coherent state with amplitude $\alpha_m$ is given. Finally, since we have a calibrated set of coherent probe states encoded in $C$ and the estimates of measurement probabilities in $R$, we can use maximum likelihood to estimate $\Pi$, which encodes the POVMs for the TESs. The corresponding results are plotted in Fig. 1 (b) in the main text. 

\subsection*{Resource counting}

\begin{figure*}[htbp]
  \centering
  \includegraphics[width=0.6\textwidth]{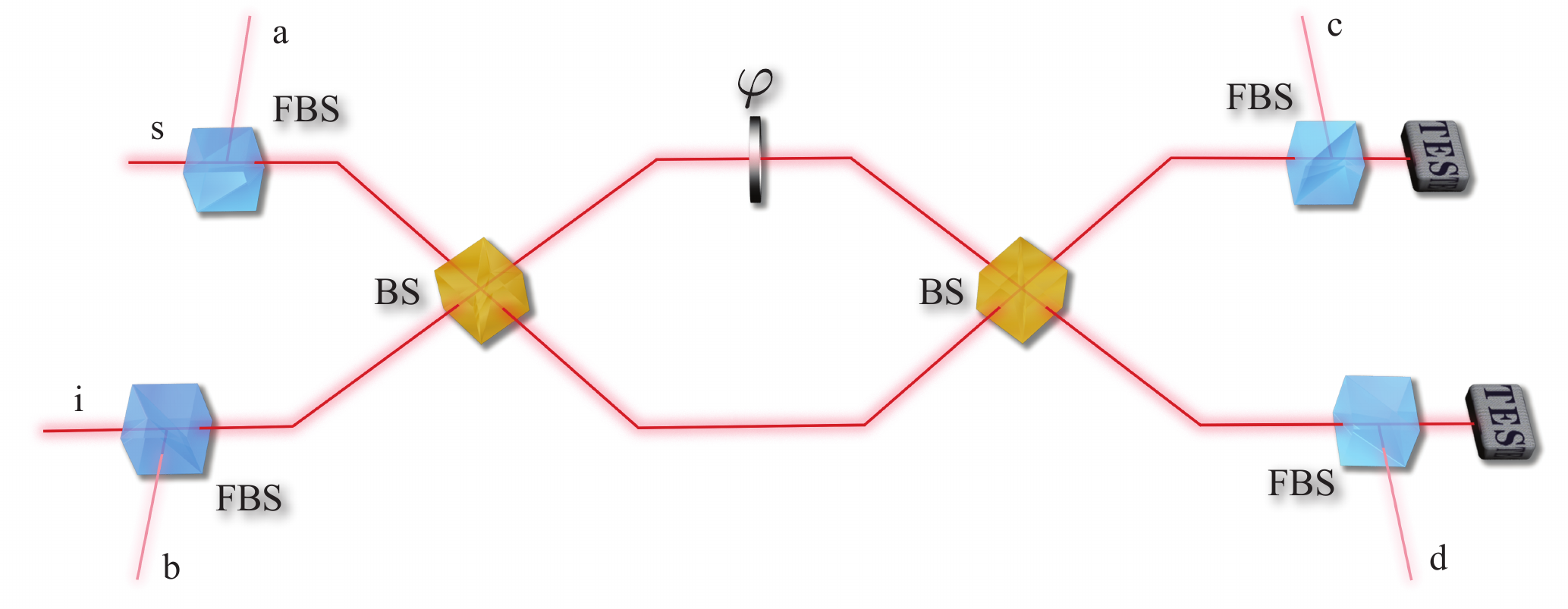}
   \caption{The theoretical model of our experimental setup when considering loss. We model the loss of the state preparation and the loss of the setup as well as fiber coupling, using several fictitious beam splitters (FBSs). In addition, we model the loss of the detector using the detector tomography data as shown in Fig. 1(b) in the main text, rather than the estimated detector efficiency.}
\label{resource}
\end{figure*}

The quantum enhancement of our metrology setup is justified by comparing the classical Fisher information of the measurement outcome with the shot-noise limit (SNL). In terms of the Fisher information, the SNL is given by $\mathcal{F}_{\text{SNL}}=\bar{n}$. To estimate $\mathcal{F}_{\text{SNL}}$, we need to accurately count the number of photons generated by the ppKTP crystal. Our common-path Mach-Zehnder interferometer can be represented schematically as a regular Mach-Zehnder interferometer as in Fig. \ref{resource}. The ppKTP crystal generates a two-mode squeezed vacuum state 
\begin{equation}
    |\psi\rangle_{1}=\sqrt{1-|z|^{2}} \sum_{n=0}^{\infty} z^{n} | n \rangle_{\text{s}} | n \rangle_{\text{i}},
\end{equation}
with $\hat{\sigma}_{1}=|\psi\rangle_{1}\langle\psi|$. 
Under realistic conditions, our model has to consider the imperfect state generation, lossy optical elements as well as the non-unity probability of free space to fiber coupling. We model these losses by adding fictitious beam splitters into the optical paths. Using the notation of $\hat{X} \circ \hat{Y} \equiv \hat{X} \hat{Y} \hat{X}^{\dagger}$, the generated two-mode squeezed vacuum state $\hat{\sigma}_{1}$ would evolve under
\begin{align}
    \hat{\sigma}_{2}&=\operatorname{Tr}_{a b}\left[\hat{U}_{a s}^{\mathrm{BS}}\left(\eta_{p} \right) \circ \hat{U}_{b i}^{\mathrm{BS}}\left(\eta_{p} \right) \circ\left(\hat{\sigma}_{1} \otimes \hat{\vartheta}_{a} \otimes \hat{\vartheta}_{b}\right)\right],\\
\hat{\sigma}_{3}&=\hat{U}_{s i}^{\mathrm{BS}}(1/2) \circ\left(\hat{P}_{s}(\theta) \otimes \hat{P}_{s}(0)\right) \circ \hat{U}_{s i}^{\mathrm{BS}}(1/2) \circ \hat{\sigma}_{2},\\
    \hat{\sigma}_{4}&=\operatorname{Tr}_{c d}\left[\hat{U}_{c s}^{\mathrm{BS}}\left(\eta_{d} \right) \circ \hat{U}_{d i}^{\mathrm{BS}}\left(\eta_{d} \right) \circ\left(\hat{\sigma}_{1} \otimes \hat{\vartheta}_{c} \otimes \hat{\vartheta}_{d}\right)\right].
\end{align}
Here $\vartheta$ is a vacuum mode, $\hat{U}_{a s}^{\mathrm{BS}}$ is the beam splitter unitary operator between mode $a$ and mode $s$ and $\hat{P}_{s}(\theta)$ is the phase shift operator on mode $s$.
And finally, the probability of detecting $j$ and $k$ photons in modes $s$ and $i$ is then given by
\begin{equation}
    p(j,k)=\sum_{m,n}\operatorname{Tr}\left[\hat{\sigma}_{4} \pi_{m}^{s}(j)\pi_{n}^{i}(k)\right],
\end{equation}
where $ \pi_{m}^{s}$ and $\pi_{n}^{i}$ are the POVM elements of the TESs. 
To estimate the average number of photons generated at the ppKTP crystal, we numerically evaluate and fit our experimental data to these equations above. Therefore, we can determine the average number of photons generated at the ppKTP crystal, using $\bar{n}=2\abs{z}^2/(1-\abs{z}^2)$. Note that due to loss during and after photon generation, $\bar{n}$ is an overestimate of the number of photons that enter the HWP that determines $\phi$, so if that number was the relevant metrological resource, $\mathcal{F}_{\text{SNL}}$ would be an overestimate of the shot-noise limit. However, since our setup is a common-path interferometer, and the spectral filter (SF) and the KTP crystal are low loss, we assume that the number of photons that pass through the HWP is close to $\bar{n}$.

It is worth highlighting the robustness of our protocol against losses. Our scheme for quantum metrology allows for phase estimation with sensitivities that surpass the shot-noise limit, even in dissipative environments characterized by 28\% loss. This prediction is obtained based on our theoretical model, by increasing losses represented by the fictitious beam splitters in each optical path. 

\subsection*{Single-photon events}

\begin{figure*}[htbp]
  \centering
  \includegraphics[width=0.45\textwidth]{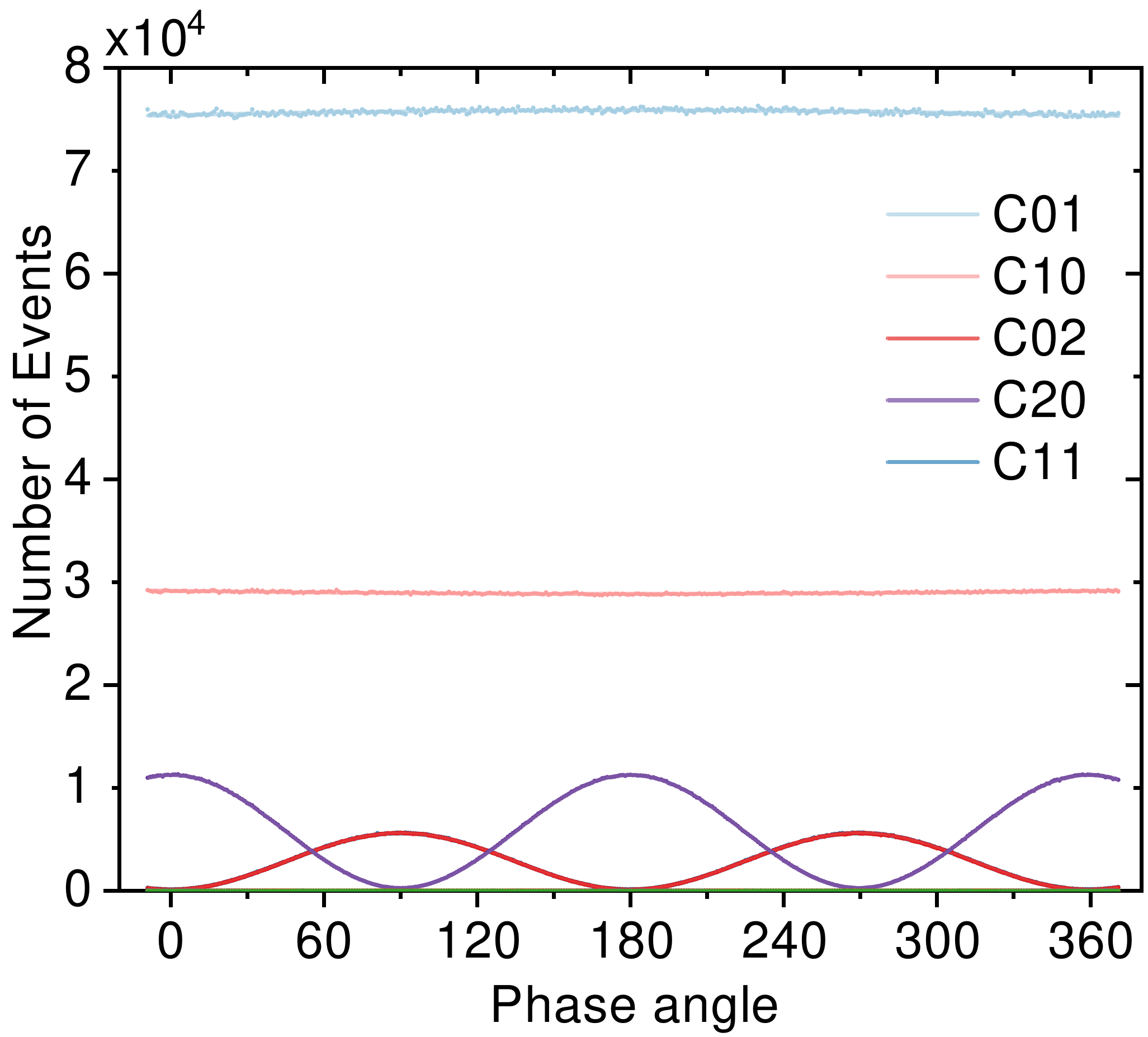}
   \caption{Experimental measurement of multiphoton interference events in our protocol.}
\label{fig5_single}
\end{figure*}

Fig. \ref{fig5_single} shows experimental multiphoton interference, which includes single-photon events. As discussed in the main text of our manuscript, due to the slightly imbalanced efficiencies of our setup, the probabilities of these single-photon events slightly vary with respect to the angle of the phase shifter. These events contribute to the Fisher information four orders of magnitude less than other terms. However, as shown in Fig. \ref{fig5_single}, the frequency of single-photon events C01 and C10 is larger than that of multiphoton events. These single-photon events are produced by two processes; photons generated by SPDC in our experiment and black-body radiation. The energy of black-body photons is similar to that of our SPDC photons. Consequently, due to the broad wavelength sensitivity and the limited energy resolution of our TESs \cite{Lita:08}, the TES signal induced by black-body photons may overlap with the signal produced by SPDC photons, leading to a ‘false’ single-photon event. This undesirable effect is negligible for two-photon coincidence events C11. Indeed, a two-photon event produced by black-body radiation is unlikely. Thus, the contamination of black-body photons dominates single-photon events C10 and C01. This is due to the high efficiency of our setup and the low pair generation probability of our SPDC source. Furthermore, we note that the magnitude of the contamination differs with respect to the detectors. We emphasize that shot-noise limit analysis presented in the body of the manuscript does not include contributions from black-body photons, since these black-body photons do not interact with the varying phase in the setup.

\end{document}